# Non-volatile reconfigurable integrated photonics enabled by broadband low-loss phase change material

*Zhuoran Fang\*, Jiajiu Zheng, Abhi Saxena, James Whitehead, Yueyang Chen, Arka Majumdar\**

Zhuoran Fang, Jiajiu Zheng, Abhi Saxena, James Whitehead, Yueyang Chen, Arka Majumdar

Department of Electrical and Computer Engineering, University of Washington, Seattle, WA 98195, USA

Arka Majumdar

Department of Physics, University of Washington, Seattle, WA 98195, USA

E-mail: rogefzr@uw.edu, arka@uw.edu

Phase change materials (PCMs) have long been used as a storage medium in rewritable compact disk and later in random access memory. In recent years, the integration of PCMs with nanophotonic structures has introduced a new paradigm for non-volatile reconfigurable optics. However, the high loss of the archetypal PCM $Ge_2Sb_2Te_5$ in both visible and telecommunication wavelengths has fundamentally limited its applications. $Sb_2S_3$ has recently emerged as a wide-bandgap PCM with transparency windows ranging from 610nm to near-IR. In this paper, the strong optical phase modulation and low optical loss of $Sb_2S_3$ are experimentally demonstrated for the first time in integrated photonic platforms at both 750nm and 1550nm. As opposed to silicon, the thermo-optic coefficient of $Sb_2S_3$ is shown to be negative, making the $Sb_2S_3$-Si hybrid platform less sensitive to thermal fluctuation. Finally, a $Sb_2S_3$ integrated non-volatile microring switch is demonstrated which can be tuned electrically between a high and low transmission state with a contrast over 30dB. Our work experimentally verified the prominent phase modification and low loss of $Sb_2S_3$ in wavelength ranges relevant for both solid-state quantum emitter and telecommunication, enabling potential applications such as optical field programmable gate array, post-fabrication trimming, and large-scale integrated quantum photonic network.

1. Introduction

Traditional means of tuning silicon photonic integrated circuits (PICs) primarily relies on thermo-optic or free carrier dispersion effect. Thermo-optic phase shifters built on silicon-on-insulator (SOI) platform have achieved device footprint <10μm[1] but are slow and power hungry (See **Table 1**). Phase shifters based on free carrier dispersion effect can significantly reduce the power consumption and increase the modulation speed but suffer in terms of large device length (>100 µm) as a result of small refractive index change (usually $\Delta n < 10^{-3}$).[2,3] To overcome these limitations, the silicon photonics community has started to head toward a hybrid approach where foreign materials are integrated with SOI waveguides to act as a tuning medium.[4] One of the promising candidates is organic electro-optic (EO) polymer which exhibits large Pockels effect and fast tuning speed.[5–7] However, these organic compounds have tendency to degrade at temperature >100°C, which renders them unsuitable to interface with the active electronics that heat up during the operation. Lithium Niobate (LN), an EO material which has long been used in free-space EO modulators, has recently gained traction thanks to the development of advanced dry etching techniques, which leads to extreme low-loss (0.25dBcm$^{-1}$) waveguides.[8] The complementary-metal-oxide-semiconductor (CMOS) compatibility of LN also makes it attractive for large-scale PIC.[9] Nevertheless, devices based of LN tend to have large footprint (>1mm) due the minimal refractive index change as a result of small Pockels coefficient. Nano-opto-electro-mechanical (NEOM) devices[10] based on plasmonic-silicon hybrid waveguide offer attojoule switching and small footprint but are relatively slow in operation. Additionally, mechanical devices often suffer from low yield and reliability issues. Finally, all the aforementioned tuning methods are volatile, necessitating a constant supply of the electric power.

Phase change materials (PCMs) can mitigate these fundamental limitations, thanks to the non-volatile phase transition and strong index modulation (typically $\Delta n > 1$), and thus can enable reconfigurable PICs for various applications including photonic switches,[11,12] photonic memory,[13] optical computing,[14] and optical neural network.[15] The suitability of PCMs for application in non-volatile reconfigurable photonics comes from its large contrast in complex refractive index upon phase transition,[16] long-term stability of multiple crystallographic phases,[17] reversible switching between the amorphous and crystalline states by both optical and electrical means,[17] cyclability of up to $10^{15}$,[18] sub-nanosecond switching speed,[19] and low switching energy.[20]

**Table 1.** Comparison of different tuning mechanisms used in integrated silicon photonics

| Tuning medium | Tuning method | Energy per bit | Speed | Insertion loss(dB) | Length of active region | Scalability | High temperature operation | Nonvolatility |
|---|---|---|---|---|---|---|---|---|
| Si[1] | Thermo-optic | 30.5nJ | ~100KHz | 0.5 | 10μm | ✓ | ✓ | ✗ |
| Si[3] | Free-carrier dispersion, PN | 4.2pJ | ~20GHz | 7.4 | 1mm | ✓ | ✓ | ✗ |
| Si[2] | Free-carrier dispersion, PIN | 5pJ | ~9GHz | 12 | 0.1mm | ✓ | ✓ | ✗ |
| Au[10] | NOEM | 130aJ | ~12MHz | 0.1 | 10μm | ✓ | ✓ | ✗ |
| EO Polymer[5,7] | Electro-optics | 0.7fJ | >100GHz | 2 | 0.5mm | ✗ | ✗ | ✗ |
| Lithium Niobate[9] | Electro-optics | 170fJ | >70GHz | 2.5 | 3mm | ✓ | ✓ | ✗ |
| GST[11,12] | Material phase change | 11nJ/86nJ [a] | ~10MHz | 1.3/35 [b] | <5um | ✓ | ✓ | ✓ |

[a] 11nJ for amorphization and 86nJ for crystallization; [b] 1.3dB for amorphous GST and 35dB for crystalline GST; The areas that GST is superior are highlighted in red.

Among various PCMs, $Ge_2Sb_2Te_5$ (GST) is by far the most studied material in integrated photonics.[21] Table 1 compares the performance metrics of a GST-based integrated photonic switch[11,12] with other popular tunning mechanisms for silicon photonics. While GST provides clear advantages over other tuning mechanisms in terms of power consumption and compact size, the performance of GST-SOI based integrated photonics is fundamentally limited by strong band to band absorption in the visible wavelengths and the near-IR (see Insertion loss in Table 1). Therefore, GST becomes impractical for large-scale PIC platforms,[22,23] and optical Field Programmable Gate Arrays (FPGAs)[24] where light is guided through numerous phase change photonic routers. While device engineering can circumvent some of the losses at 1550nm[25], this comes at the expense of increased device length, and such approach does not work near visible wavelength, where the optical loss is very large. A PCM with wide bandgap can overcome these limitations. In this paper, we investigate a wide bandgap PCM $Sb_2S_3$ which exhibits broadband transparency from 610nm to near-IR.[26] The strong optical phase modulation and low optical loss of $Sb_2S_3$ are experimentally demonstrated for the first time on two different integrated photonic platforms: Silicon Nitride (SiN) and SOI at 750nm and 1550nm, respectively. As opposed to silicon, the thermo-optic coefficient of $Sb_2S_3$ is shown to be strongly negative, making the $Sb_2S_3$-Si hybrid platform less sensitive to thermal fluctuation. Finally, a $Sb_2S_3$ integrated non-volatile microring switch is demonstrated which can be tuned electrically between a high and low transmission state with an extinction ratio of over 30dB. The work experimentally verified the prominent optical phase modification and low loss of $Sb_2S_3$ (abbreviated as SbS in the rest of the paper) in a broad wavelength range, enabling potential applications such as optical FPGAs,[24] post-fabrication trimming,[27] and large-scale integrated quantum photonic networks.[28]

## 2. Results and Discussion

2.1 Materials Characterization:

Four samples of 20nm-thick SbS were deposited by sputtering (see **Experimental Section**) on silicon substrates and were later annealed for 20mins at four different temperatures, ranging from 423K to 573K. The optical constants of SbS in amorphous and crystalline states were first measured by ellipsometry (**Figure 1a**) for optical mode simulation. The phase transition is confirmed in both X-ray diffraction (XRD) (Figure 1b) and Raman spectroscopy (Figure 1c). We found that the material started to crystallize at ~523K and continued to crystallize through 573K, as seen from the emerging XRD peaks in Figure 1b. The Bragg diffraction angles $2\theta$ that give rise to constructive interference match with the previous literatures[29–32] and correspond to the characteristic crystallographic planes in typical SbS crystals. The optical micrographs (right panel of Figure 1b) support the XRD results, showing the nucleation and growth of SbS polycrystalline domains with increasing temperature. The grain growth was incomplete at 523K and reached a completion at 573K. This is further corroborated by the Raman spectrum showing the characteristic Raman shift in both as-deposited amorphous and annealed crystalline SbS, denoted as aSbS and cSbS respectively, with good match to the previous literatures.[30,32]

**Table 2.** Comparison of $\Delta n$ and $k_c$ of SbS, GST, and GSST at 633nm and 1550nm. $k_c$ (extinction coefficient) is quoted for the crystalline state of the PCM. $\Delta n$ is the refractive index change from amorphous to crystalline state.

|  | $\Delta n$ at 633nm | $k_c$ at 633nm | $\Delta n/k_c$ at 633nm | $\Delta n$ at 1550nm | $k_c$ at 1550nm | $\Delta n/k_c$ at 1550nm |
|---|---|---|---|---|---|---|
| GST[11] | 0.16 | 3.82 | 0.04 | 2.74 | 1.09 | 2.51 |
| GSST[33] | 0.62 | 2.56 | 0.24 | 1.75 | 0.42 | 4.17 |
| SbS | 0.87 | 0.56 | 1.55 | 0.54 | 0.05 | 10.8 |

The minimal optical loss of SbS originates from its wide bandgap[26] (~1.7-2eV), which leads to suppressed band to band absorption from 610nm. Even in its crystalline state, SbS exhibits low loss ($k$~0.05) at 1550nm. This is superior to the recently-reported broadband transparent GSST[33–35] which has near-zero loss at 1550nm in amorphous state but suffers from non-negligible loss in the crystalline state (see **Table 2**). One potential limitation for SbS, however, is the simultaneous reduction in both $\Delta n$ and $k$, as a result of Kramers-Kronig relation, and hence smaller phase shift per unit length. However, this could be easily overcome by using longer or thicker SbS (as discussed later in the paper), thanks to the reduced optical absorption.

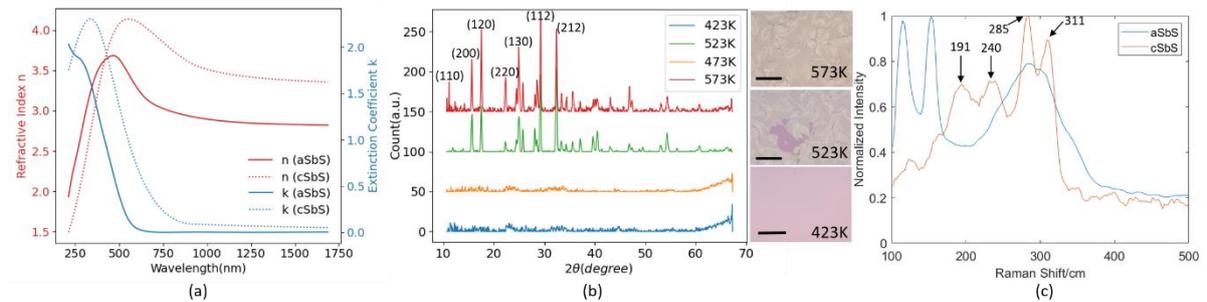

(a) (b) (c)

**Figure 1.** Material characterization of SbS. (a) Fit refractive index (*n*) and extinction coefficient (*k*) of SbS in amorphous and crystalline states. (b) X-ray diffraction spectrum of SbS annealed at four different temperatures for 20mins. Characteristic diffraction peaks are indexed based on literatures. The right panel shows the morphological evolution of SbS polycrystal grains as the annealing temperature increases (scale bar is 200µm). (c) Raman spectrum of amorphous and crystalline (annealed at 573K) SbS. The characteristic Raman shifts of cSbS are indexed according to the literature.

The effective index and mode profiles of the waveguides with different thicknesses of SbS on SOI waveguide are simulated using Lumerical Mode Solutions to find the optimal thickness. Figures 2a and 2b show that the fundamental quasi-transverse electric (TE) mode at 1550nm is confined in the waveguide and there is a large effective index change ($\Delta n_{eff} \sim 0.04$) due to the phase transition of 20nm SbS. Thanks to the low loss of SbS, we get a large ratio between the effective index change and the loss at the crystalline state ($\frac{\Delta n_{eff}}{k_{eff}} \sim 8.62$). Note that the final capping layer of 10nm SiN is not considered in the reported simulation. The extra SiN capping raises the effective index by 0.06 but does not change the $\Delta n_{eff}$ and loss. The increased imaginary part in cSbS indicates the increase of optical absorption in the crystalline state. We also plot the change of effective index and optical absorption as a function of SbS thickness (Figure 2e). The index contrast increases with increasing PCM thickness (as the optical mode has stronger interaction with the PCM) but the optical absorption also becomes larger. This is illustrated in Figures 2c and 2d with 66nm of SbS integrated on silicon waveguide ($\Delta n_{eff} \sim 0.07\ and\ \frac{\Delta n_{eff}}{k_{eff}} \sim 7.65$). Based on our simulation, we chose the SbS thickness as 20nm to reduce the excess loss ($\alpha = \frac{4\pi k_{eff}}{\lambda} = 0.19 dB\mu m^{-1}\ and\ \Delta n_{eff} \sim 0.04$). The index contrast far surpasses the typical Si thermal phase shifter which can only achieve $\Delta n_{eff} < 1.8 \times 10^{-4}$ per Kelvin change in temperature.[36] The 20nm thickness is also chosen to facilitate the thermal diffusion as the PCM needs to be quenched rapidly during amorphization so thinner film has faster cooling rate.[37]

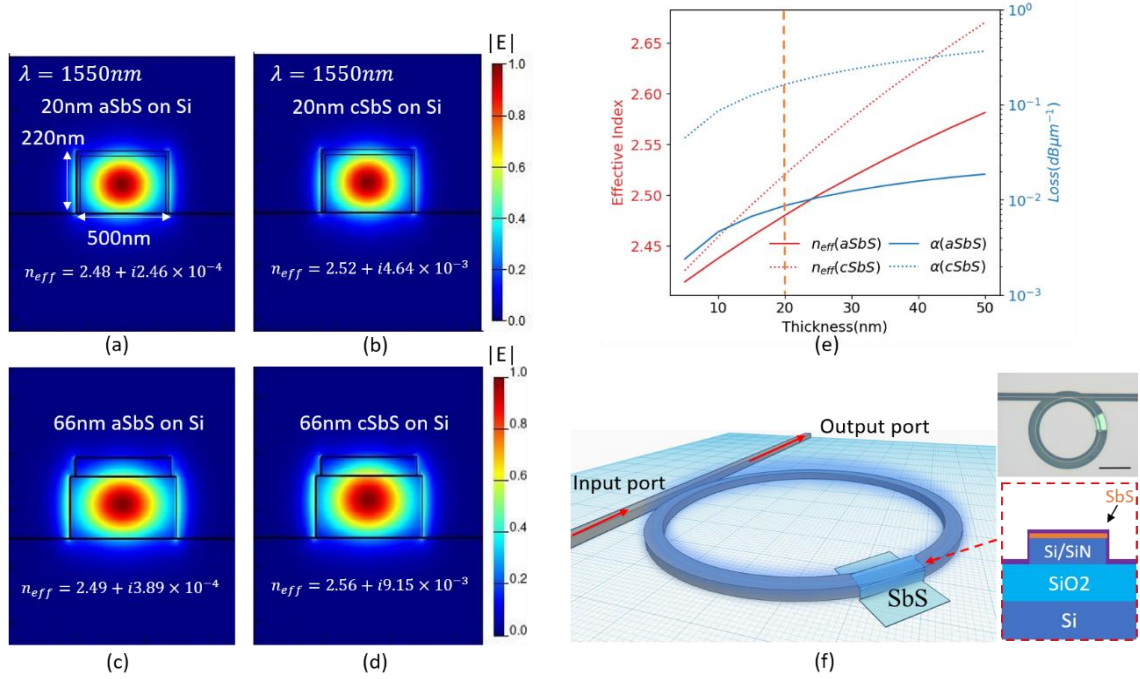

**Figure 2.** Optical mode simulation of SbS on SOI. (a), (b) Simulated fundamental TE mode profiles of 20nm aSbS/cSbS on SOI hybrid waveguide at 1550nm using the optical constants extracted from ellipsometry. Note here that the waveguide sidewalls are also covered with SbS (to be consistent with the experiments reported later in the paper). (c), (d) Simulated fundamental quasi-TE mode profiles of 66nm aSbS/cSbS on SOI hybrid waveguide at 1550nm. Here, SbS is only placed on top of the waveguide to be consistent with the experiment. (e) Variation of effective index and absorption coefficient with the SbS thickness. (f) Schematic of microring resonators with SbS patterned on the ring. The inset shows optical micrograph (scale bar: 20µm) of SbS-capped ring and the cross-section of the device.

2.2 SbS on Si microring resonators:

Varying lengths of 20nm and 66nm thick SbS are deposited on Si microring resonators of radius 20 µm (device schematic shown in Figure 2f) using standard fabrication techniques (See Experimental Section). The insets show the optical micrograph of Si microring patterned with 10µm-long, 20nm-thick SbS and the cross-section of the device. The PCM is capped with 10nm of PECVD SiN to prevent oxidation during annealing.

The phase shift induced by GST[11] and different thicknesses of SbS near 1550nm is compared in **Figure 3a** where the length of PCM is fixed at 10µm. It can be seen that due to the high loss of cGST ($\alpha = 7.6 dB\mu m^{-1}$), the resonance is completely destroyed at crystalline state, whereas the resonance is clearly visible for the 20nm and 66nm thick cSbS capped rings, implying that cSbS has much lower optical loss than GST. The spectral shift caused by phase transition of 66nm SbS is 0.947nm compared to 0.53nm for 20nm SbS, which matches the 75% increase in $\Delta n_{eff}$ from the mode simulation (Figures 2a-2d). It is also worth pointing out that the microring is designed to be critically coupled at crystalline state which is the reason why the extinction ratio of the resonance increases dramatically as a result of the enhanced loss due to the phase transition. Figure 3b shows how the spectral shift and extinction ratio evolved as the length of 66nm-thick SbS increases from 0 to 50µm across the telecommunication C-band. The red shift becomes increasingly large such that the resonance eventually shifts over one free spectral range of the microring at 50µm. The rise in optical loss with length is confirmed by the transition from overcoupling (small resonance dip) at 0µm, near critical coupling (maximum dip) at 10µm, and strong undercoupling (small dip) at 50µm. Notice that a red shift (~0.8nm) is seen on the 0µm one (i.e., no SbS) due to a global capping of PECVD SiN before thermal annealing, which does not affect our conclusion as the capping layer is lossless and affects all the devices equally.

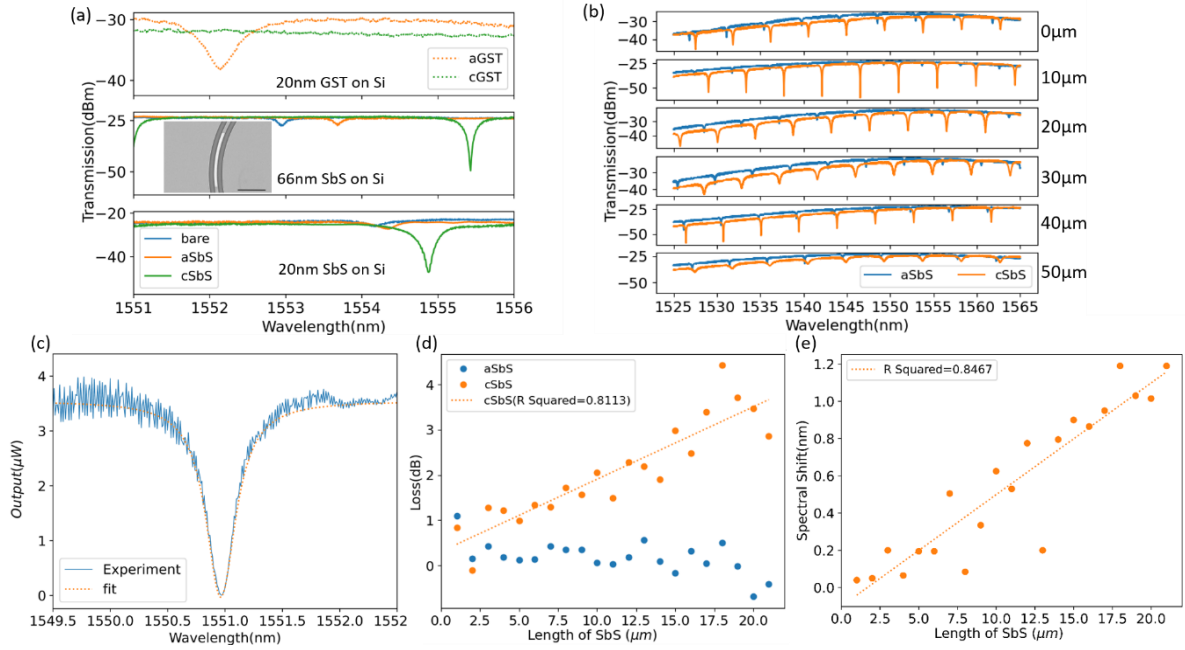

**Figure 3.** SbS on Si microring resonators. (a) Optical spectrum of microring resonators capped with 20nm GST(top), 66nm SbS(middle), and 20nm SbS(bottom) in amorphous and crystalline states. The length of PCM was kept at 10μm. The inset of middle plot shows the SEM of 66nm SbS placed only on top of the waveguide. (b) Transmission of ring resonators over the C band for increasing length of 66nm aSbS and cSbS on SOI. (c) Lorentzian fitting of measured resonance dip of $Q$ factor ~ 5000. (d) Optical loss of 20nm thick aSbS and cSbS on SOI at various lengths. (e) Spectral shift from amorphous to crystalline with increasing length of 20nm thick SbS on SOI.

To analyze the phase modulation effect and loss quantitatively, the resonance dip near 1550nm is fit with a Lorentzian function to extract the quality factor ($Q$) and the loss (Figure 3c). The optical phase modulation effect is quantified as the spectral shift $\Delta\lambda$ from aSbS to cSbS, while the attenuation is extracted from the $Q$ factor reduction from amorphous to crystalline state. The results are plotted in Figures 3d-3e. The experimental data show a good linear fit where the red shift due to the SbS phase transition increases linearly with the length of SbS. The spectral shift per unit length of SbS is extracted to be 0.060±0.006 nm·µm$^{-1}$. The loss introduced by SbS is estimated from the $Q$ factors using the relation[11]:

$$Loss = 2\pi R \frac{2\pi n_g}{\lambda_0}\left(\frac{1}{Q}-\frac{1}{Q_0}\right) = \frac{2\pi\lambda_0}{FSR}\left(\frac{1}{Q}-\frac{1}{Q_0}\right) \quad (1)$$

where $n_g$ is the group index, $Q$ is the quality factor of the ring after SbS phase transition and $Q_0$ is the $Q$ factor of the ring with as-deposited SbS, $FSR$ is the free spectral range. The loss of cSbS (orange line) increases linearly with the PCM length whereas the aSbS (blue line) remains almost flat, signifying extremely low loss as also confirmed by the mode simulation (Figure 2a). The attenuation of cSbS is estimated to be 0.16±0.02dBµm$^{-1}$, almost 50 times smaller than that of cGST.[11]

The spectral shift induced by PCM on ring can be numerically calculated using equation[11]:

$$\frac{\Delta\lambda}{L_{SbS}} = \frac{\Delta n_{eff}\lambda_0}{2\pi R n_{eff0}} \quad (2)$$

where $R$ is the radius of the microring, $\lambda_0$ $and$ $n_{eff0}$ are the resonance wavelength and effective index before phase change, respectively. $\Delta n_{eff}$ due to SbS phase change is extracted from the mode simulation as discussed above. The simulation results are compared with the experimental results in **Table 3**.

**Table 3.** Comparison between the simulation and experimental results

|  | Simulation | Experiment |
| --- | --- | --- |
| $\Delta\lambda(nm/\mu m)$ | 0.194 | 0.060 ± 0.006 |
| $\Delta\alpha(dB/\mu m)$ | 0.17 | 0.16 ± 0.02 |

The extracted $\Delta n_{eff}$ from experiment is 0.012 compared to 0.04 from simulation (which uses the ellipsometry data from blanket SbS measurement). This reflects a potential deviation of material optical properties, especially the real part of refractive index, as the PCM shrinks down from 50nm-thick blanket film (used for ellipsometry) to thin nanoscale patches on waveguides. Such variation is however geometry and thickness dependent, as found later in the paper that for a thicker SbS film patterned only on top of the waveguides (e.g., SbS on SiN and rib SOI waveguide) the discrepancy becomes smaller. A possible explanation for this is the nonuniform coverage of SbS on the waveguide sidewall – SbS deposited on the sidewall is likely to be thinner than on the top of the waveguide due to the slight directional nature of the sputtering. On the other hand, the ellipsometry does predict the imaginary part of refractive index accurately which implies that the material's optical loss does not vary significantly as the dimensions reduce. This result provides an insight to the design of PCM based photonic devices where the discrepancy between nano-patterned and the bulk material's optical properties must be considered.

2.3 Electrical Switching of SbS:

To demonstrate the feasibility of an active integrated device based on SbS, an ITO external heater is patterned on top of the PCM to achieve on-chip electrical actuation of SbS phase transition (see Experimental Section for fabrication details and the optical characterization setup), as shown in **Figure 4a**. By applying voltage pulses across the ITO, electrical current causes Joule heating and subsequently phase transition of the PCM. Here, a microring modulator design[38] is adopted to amplify the modulation effect of SbS. The Si waveguide is partially etched by 140nm, as opposed to fully etch, to leave an 80nm-thick Si slab which acts as a heat sink to facilitate thermal diffusion from the PCM. It also helps the lift-off of ITO film as thinner ITO can be used to ensure conformal step coverage. The false color SEM in Figure 4b shows fabricated device where the Si waveguide is conformally covered with ITO. The SbS is deposited beneath the ITO, and hence is not visible in the SEM. Figures 4c-4d demonstrate the electrical switching of 8μm-long SbS into its crystalline state by a DC voltage sweep from 0 to 1V at power of 13mW. The I-V curve in the inset of Figure 4d shows a relatively linear relationship – a proof that ITO has stable resistance over this voltage/temperature range. Heat transfer simulation[37] (**Figure S1** in **Supporting information**) also confirms that the PCM layer has reached a temperature of 650K, well above the crystallization temperature of SbS (573K) but below the melting point (823K). The DC sweep allows extended period for long range atomic diffusion, during which the atoms can settle in the most energetically favourably positions. The red shift is a result of PCM phase transition because it was found that the refractive index of amorphous ITO remains relatively constant upon heating up to 773K[39] while the thermo-optic effect is volatile. The spectral shift of 0.4nm is slightly smaller than what is predicted by the simulation of rib waveguide (0.456nm, see Figure S1 Supporting information) and measured from fully etched waveguide (0.48nm) due to the refractive index deviation (as found earlier) and less mode confinement. However, since the ring is designed to be critically coupled at crystalline state, it is still large enough to give an extinction ratio of over 30dB in optical transmission with minimal increase in the resonance linewidth. The strong phase modulation at a cost of small loss shows that SbS is a promising tuning medium for next-generation non-volatile large-scale PICs.

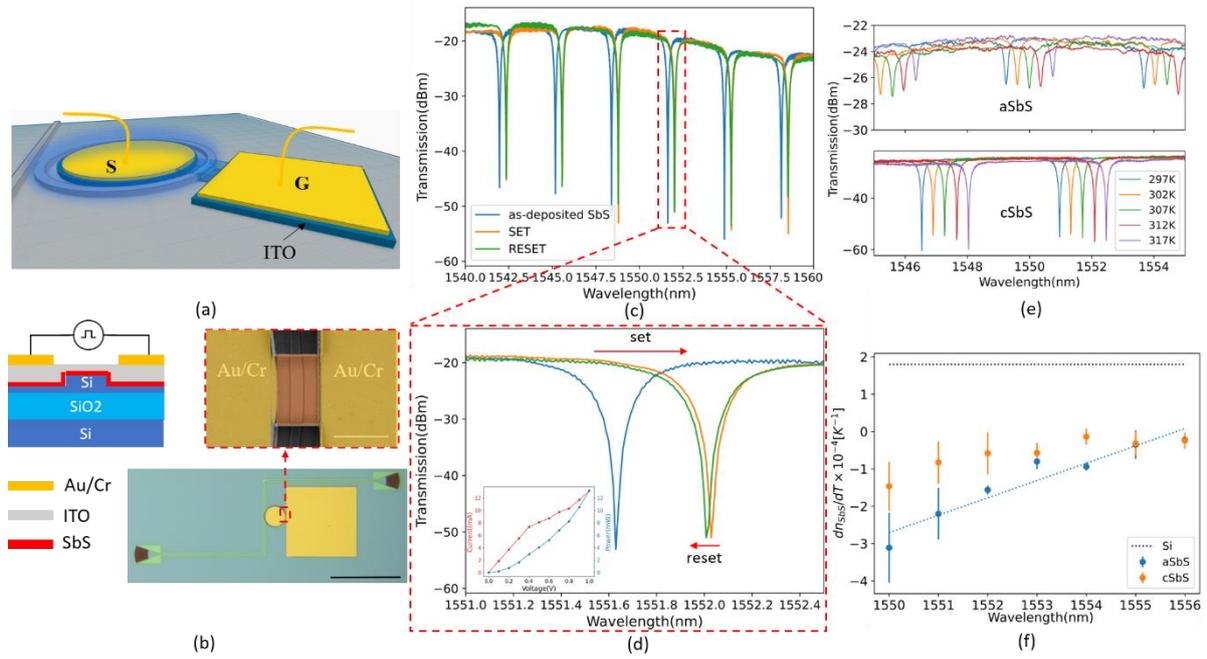

**Figure 4.** Electrical Switching and thermo-optic effect of SbS. (a) Schematic of integrated photonic microring switch based on SbS. (b) Cross-section of the device, optical micrograph (scale bar: 200µm) and false color SEM (scale bar: 3µm) of the fabricated device. (c), (d) Transmission spectrum showing the SET and RESET operation as red and blue shift of the resonance dip. The inset of (d) shows the I-V curve (red) and power(blue) during a 0-1V DC sweep that triggered the SET operation. (e) Spectral shift caused by raising the temperature from 297K to 317K of microring capped with 10µm-long, 66nm-thick SbS. (f) Extracted thermo-optic coefficient of aSbS and cSbS compared with silicon.

The device can be reset by applying a 6V 200ns square pulse using a function generator, causing a slight blue shift. Heat transfer simulation (Figure S1 Supporting information) shows that the temperature is raised well above the melting point of SbS (823K) and below the melting point of Si waveguide. The trailing edge of the pulse was set to be 8ns to allow the melt-quench of the PCM during which the fast-moving atoms are quickly frozen into a disordered state. However, the RESET spectrum does not perfectly overlap with the original spectrum while the *Q* factor remains roughly the same as the SET spectrum. We speculate that the abrupt resistivity increase of ITO at 650K[40] gives rise to regional thermal hot spots that causes nonuniform heating of the PCM. Pulses of higher voltages led to further blue shift and broadening of resonance dip, which implies that the PCM amorphization is accompanied by waveguide damage, as some local hotspots may have temperature exceeding the melting point of Si. Transparent conductors whose conductivity is thermally stable such as graphene[41] and FTO[42] could provide a better solution for operation with large number of cycles.

2.4 Thermo-optic Effect of SbS:

Due to the large positive thermo-optic (TO) effect of silicon,[36] SOI platform is prone to thermal fluctuations, particularly for high-*Q* optical resonators.[43] Integrated metal heaters near the resonators are normally used to stabilize the temperature and prevent resonance drift from the temperature variation in the environment.[44,45] Here, we showed that SbS exhibits strong negative TO effect that counteracts the positive TO effect of silicon. The overall TO coefficient of SbS-SOI hybrid waveguide is hence lower than pure Si waveguide, making the hybrid system more thermally stable. By heating the ring resonators capped with different lengths of SbS, the TO coefficient of SbS on SOI is extracted in both amorphous and crystalline states near 1550nm. Thicker SbS (66nm) is used to amplify the minimal change in resonance spectrum from the TO effect of SbS. Note that the PCM is only placed on top of the waveguide to simplify the calculation of mode confinement factor [Figures 2c and 2d]. The TO coefficient of SbS-SOI hybrid waveguide can be approximated to the first order as[46]:

$$\frac{dn_{eff}}{dT} = \Gamma_{Si}(\lambda)\frac{dn_{Si}}{dT}(\lambda) + \Gamma_{SbS}\frac{dn_{SbS}}{dT}(\lambda) \qquad (3)$$

where the mode confinement factors $\Gamma$[47] of Si waveguide and SbS are wavelength dependent and can be calculated using Lumerical Mode Solutions (see S2 in Supporting information). By adjusting the temperature of a thermal electric controller beneath the chip, we can tune the resonances (Figure 4e). Using the resonance shift and the temperature change, the collective TO coefficient of SbS-Si hybrid waveguide can be expressed as (see S2 in for derivation):

$$\frac{dn_{eff}}{dT} = \left(\frac{\lambda_{res}n_{eff}^{Si}}{\lambda_{res0}\Delta T} - \frac{n_{eff}^{Si}}{dT} - \frac{dn_{eff}^{Si}}{dT}\right)\frac{2\pi R - L_{SbS}}{L_{SbS}} + \left(\frac{\lambda_{res}}{\lambda_{res0}} - 1\right)\frac{n_{eff0}}{\Delta T} \quad (4)$$

Note that $\frac{dn_{eff}}{dT}$ and $\frac{dn_{eff}^{Si}}{dT}$ in Equation (4) not only describe the TO effect of SbS-SOI hybrid waveguide and Si waveguide, but also include the effect of dispersion which has to be subtracted to extract the pure temperature effect to the refractive index. $n_{eff0}$ and $n_{eff}^{Si}$ are the effective indices of SbS-Si hybrid waveguide and SOI waveguide respectively at the resonance wavelength before the temperature change of $\Delta T$. $\lambda_{res0}$ and $\lambda_{res}$ are the resonance wavelengths before and after the temperature change, respectively. Substituting (3) into (4) hence yields the expression of TO coefficient of pristine SbS in terms of the measurable parameters $\lambda_{res0}$, $\lambda_{res}$, and $\Delta T$. Here, the length of the SbS is varied from 0 to 80µm (nine rings) and each device is measured at 5 different temperatures (i.e., for four $\Delta T$s) from 297K to 317K, giving four values of TO coefficient for each length of SbS. No definite dependency of SbS TO coefficient on temperature is observed in the experiment. Hence, it is assumed that the change in the TO coefficient with temperature is negligible for both Si and SbS in this temperature range. Figure 4f shows the wavelength dependent TO coefficients of both amorphous and crystalline SbS compared with Si near 1550nm.[36] The data points are calculated as the average of TO coefficients across four $\Delta T$s while the error bars measure the standard deviation from the average of four. The TO coefficient of aSbS exhibits a strong linear dependency to wavelength, whereas the TO coefficient of cSbS is relatively wavelength independent. The TO coefficient of aSbS and cSbS at 1550nm are estimated to be $-3.11 \times 10^{-4} K^{-1}$ and $-7.28 \times 10^{-5} K^{-1}$. The strong negative TO effect of SbS shows that the SbS-SOI hybrid platform has less sensitivity to heat fluctuation than SOI does and is suitable for use in non-volatile integrated photonics.

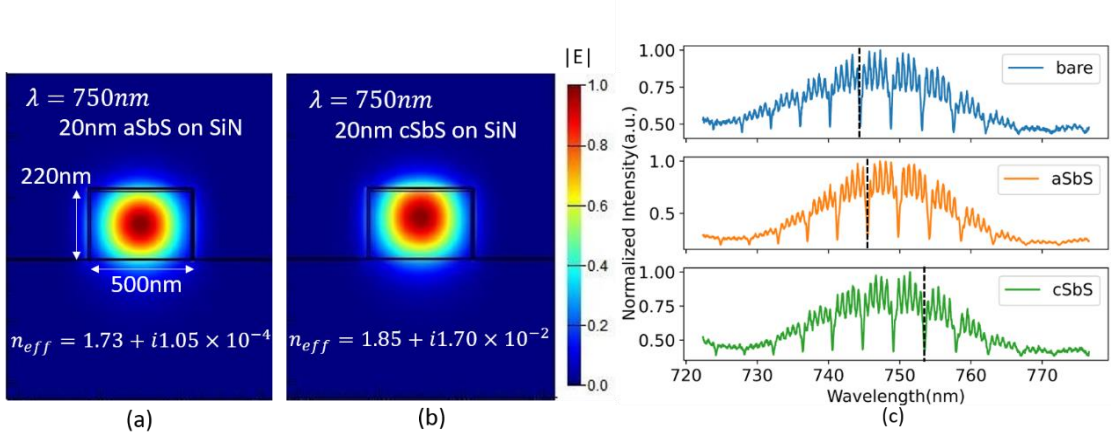

**Figure 5.** SbS on SiN microrings. (a) Simulated fundamental quasi-TE mode profiles of 20nm aSbS/cSbS on SiN hybrid waveguide at 750nm. Note here that the SbS is only place on top of the waveguide. (b) Transmission spectrum of SiN ring resonators capped with 1μm aSbS and cSbS near 750nm. The dotted lines indicate the same resonance.

2.5 SbS on SiN microring resonators:

Finally, since SbS is transparent near visible wavelengths, the tuning ability of SbS on SiN platform is also investigated. Specifically, we probed the wavelength regime near 750nm, where quantum emitters like SiV in diamond[48] and quantum materials like $WSe_2$ emit.[49] A second chip was fabricated (See Experimental Section) on SiN-on-oxide platform with similar design to probe the optical properties of SbS near 750nm. The wide bandgap of SiN makes it transparent in the visible wavelengths and is widely used in integrated quantum optics.[46,50]

**Figures 5a-b** show the mode simulation of 20nm SbS on SiN at 750nm wavelength where the effective index change $\Delta n_{eff} = 1.2$ is three times that of SbS on SOI at 1550nm. Figure 5c shows the transmission spectrum of 1µm-long SbS on SiN ring (10µm radius) before and after thermal annealing. Owing to low optical losses, the resonances are clearly visible in both amorphous and crystalline states from 720nm to 770nm. The dotted line in the top plot indicates the original resonance position and the dotted lines in the plots below indicate the new resonance positions. A red shift of 8nm can be seen from aSbS to cSbS versus the predicted 8.23nm·µm$^{-1}$ shift based on $\Delta n_{eff} = 1.2$ and equation (2). The phase modulation effect of SbS near 750nm is much more prominent than the telecom C-band, demonstrating its unlimited potential to be used in tunable visible integrated photonics.

3. Conclusion

In this paper, an emerging wide-bandgap phase change material $Sb_2S_3$ is investigated for its applications in non-volatile reconfigurable photonics. The XRD and Raman data have confirmed the phase transition of $Sb_2S_3$ from amorphous to crystalline state while ellipsometry shows that the material exhibits broadband transparency from 610nm to near-IR. The large refractive index contrast and low absorptive loss of $Sb_2S_3$ are verified experimentally for the first time on SOI and SiN integrated photonics platform using microring resonators at the telecom C-band and near visible wavelengths. The measured resonance shifts are 0.06nm·µm$^{-1}$ at 1550nm and 8nm·µm$^{-1}$ at 750nm while the optical losses in crystalline state are 0.16dBµm$^{-1}$ at 1550nm and zero in amorphous sates. An on-chip non-volatile $Sb_2S_3$-SOI microring switch with low-loss, large optical phase shift ($\Delta\lambda\sim0.4nm$) and extinction ratio(>30dB) is demonstrated. Additionally, the thermo-optic coefficient of aSbS and cSbS at 1550nm is extracted to be $-3.11 \times 10^{-4} K^{-1}\ and - 7.28 \times 10^{-5} K^{-1}$ respectively, with opposite sign to that of Si, showing that the SbS-SOI hybrid waveguide is more stable to temperature variation than Si waveguide. Potential future work will aim to show the electrical tunability of $Sb_2S_3$ on integrated SiN platform near 750nm as well as large endurance of $Sb_2S_3$ integrated photonic switches. Our experiments introduce $Sb_2S_3$ as a promising candidate for the low loss non-volatile tuning of large-scale PIC at lower energy consumption and smaller footprint and will find wide applications in fields such as optical FPGAs, post-fabrication trimming, and non-volatile quantum integrated optics.

4. **Experimental Section**

*SOI Device Fabrication:* The SbS-on-SOI microring resonators were fabricated on a 220-nm thick silicon layer on top of a 3-μm-thick buried oxide layer (SOITECH). The pattern was defined by a JEOLJBX-6300FS 100kV electron-beam lithography (EBL) system using positive tone ZEP-520A resist. 220 nm fully etched ridge waveguides were made by an inductively coupled plasma reactive ion etching (ICP-RIE) process. A second EBL exposure using positive tone poly(methyl methacrylate) (PMMA) resist was subsequently carried out to create windows for the SbS deposition. After development, 20nm SbS was DC sputtered from a SbS target (Plasmaterial Ltd.) in a magnetron Sputter System (Lesker Lab18) under Ar atmosphere at 27W and base pressure of ~$5 \times 10^{-7}$ Torr. The Ar flow rate is controlled to maintain a sputtering pressure of 3.5mTorr. The plasma is stabilized for 5 minutes before the actual deposition. The lift-off of SbS was completed using methylene chloride followed by a global capping of 10nm PECVD SiN at 125°C to prevent oxidation. Note that the SbS samples used for ellipsometry, Raman and XRD were deposited on Si [100] wafers and capped with 10nm sputtered ITO. The SbS-ITO integrated photonic switch were fabricated in a similar process where the waveguide was only partially etched by 140nm for easier liftoff of ITO. A 200nm thick ITO layer was sputtered following the 20nm SbS deposition without breaking vacuum. After the lift-off of the SbS/ITO layers, an extra EBL overlay and electronbeam evaporation of Au/Cr (60nm/30nm) were done to pattern the electrode pads. For crystallization of SbS, rapid thermal annealing (RTA) at 300 °C for 20 min was performed under N2 atmosphere.

*SiN device fabrication:* The SiN microring resonators were fabricated on a 220-nm-thick SiN membrane grown via LPCVD on 4 μm of thermal oxide on silicon (Rogue Valley Microelectronics). We spun roughly 400 nm of ZEP520A, which was coated with a thin layer of Pt/Au that served as a charging layer. The resist was then patterned by EBL and the pattern was transferred to the SiN using the same RIE in $CHF_3/O_2$ chemistry. The SbS was deposited onto the SiN rings using the same process mentioned above.

*Experimental setup:* The microrings on SOI were characterized by a vertical fiber-coupling setup.[12] All the measurements were performed under ambient conditions while the temperature of the stage was fixed at ~24°C by a thermoelectric controller (TEC, TE Technology TC-720) to prohibit the serious thermal shift of the resonators. The input light was provided by a tunable continuous-wave laser (Santec TSL-510) and its polarization was controlled by a manual fiber polarization controller (Thorlabs FPC526) to match the fundamental quasi-TE mode of the waveguides.

A low-noise power meter (Keysight 81634B) was used to collect the static optical output from the grating couplers. The transmission spectrum measurement was performed after the fabrication of bare ring resonators, deposition of SbS, and annealing respectively to extract the spectral shift and change in optical loss in each step. For the electrical characterization of SbS-ITO photonic switch, the electrical signals were applied to the metal contacts by a pair of DC probes controlled by two probe positioners (Cascade Microtech DPP105-M-AI-S). The current sweep and voltage measurement were provided by a source meter (Keithley 2450) and the Reset pulses were generated from a pulse function arbitrary generator (Keysight 81160A).

The SiN microrings were characterized via measuring the transmission using a confocal microscopy setup.[46] A super continuum light source (Fianium WhiteLase Micro) is focused on the grating coupler through the objective lens, and a moveable pinhole is used to pick up only the signal coming out from the other grating and then send it to a spectrometer. The spectrometer is equipped with a Princeton Instruments PIXIS CCD with an IsoPlane SCT-320 Imaging Spectrograph.

**Acknowledgements**

Z.F. and A.M. conceived the project. Z.F. simulated, designed, and fabricated the devices. Z. F. performed the experiments. J.Z. helped with the experiments and simulations. A.S. performed the SiN optical measurements. J.W. helped with the SEM imaging. Y.C. advised on the SiN fabrication. A.M. supervised the overall progress of the project. Z.F. wrote the manuscript with input from all the authors.

The research is funded by National Science Foundation (NSF-1640986, NSF-2003509), and ONR-YIP Award. Part of this work was conducted at the Washington Nanofabrication Facility / Molecular Analysis Facility, a National Nanotechnology Coordinated Infrastructure (NNCI) site at the University of Washington, which is supported in part by funds from the National Science Foundation (awards NNCI-1542101, 1337840 and 0335765), the National Institutes of Health, the Molecular Engineering & Sciences Institute, the Clean Energy Institute, the Washington Research Foundation, the M. J. Murdock Charitable Trust, Altatech, ClassOne Technology, GCE Market, Google, and SPTS.